\documentclass{aa}  

\def\mue{\mbox{$\mu_{\rm e}$}}
\def\re{\mbox{$r_{\rm e}$}}
\def\ha{\mbox{H$\alpha$}}

\def\Kelv{$K$}
\def\micron{\mbox{$\mu$m}}
\def\mm1{mm$^{-1}$}
\def\kms{\rm km~s$^{-1}$}
\def\kmsM{\rm km~s$^{-1}~Mpc^{-1}$}
\def\mga{\rm mag~arcsec$^{-2}$}

\def\r25{$r_{25}$}

\def\gradip{\hbox{\rlap{\hbox{.}}\raise 5.truept \hbox{{\small $\circ$}}}}
\def\mincir{\ \raise-2.truept\hbox{\rlap{\hbox{$\sim$}}\raise5.truept
    \hbox{$<$}\ }}
\def\magcir{\ \raise-2.truept\hbox{\rlap{\hbox{$\sim$}}\raise5.truept
    \hbox{$>$}\ }}
\def\etal{et al.\/}
\def\cf{{\it cf.\/}}
\def\ie{{\it i.e.\/}}
\def\eg{{\it e.g.\/}}

\usepackage{times}
\usepackage{mathptm}
\fontfamily{ptmcm}\selectfont
\usepackage{graphicx}

\begin{document}
   \thesaurus{04         
	      (11.09.1 NGC~128;  
	       11.05.1;
               11.11.1;   
	       11.16.1;
	       11.19.6;
	       11.07.1)}
%
   \title{Structure and kinematics of the peculiar galaxy NGC~128.
   \thanks{Based on observations taken at ESO La Silla and at the
           Special Astrophysical Observatory (SAO) of the Russian 
           Academy of Sciences (RAS).}}
   \author{M. D'Onofrio \inst{1}
   \and M. Capaccioli \inst{2,3} 
   \and P. Merluzzi \inst{2}
   \and S. Zaggia \inst{2}
   \and J. Boulesteix \inst{4}}
   \offprints{M. D'Onofrio}
   \institute{Dipartimento di Astronomia, Universit\`a\ di Padova, vicolo
              dell'Osservatorio 5, I-35122 Padova, Italy
\and
              Osservatorio Astronomico di Capodimonte, via Moiariello 16,
              I-80131 Napoli, Italy
\and
              Dipartimento di Scienze Fisiche, Universit\`a\ di Napoli,
              Mostra d'Oltremare, Padiglione 19, I-80125, Napoli, Italy 
\and
              Observatoire de Marseille, 2 Place Le Verrier, F-13248 Marseille
              Cedex 04, France}
 
   \date{Received ; accepted }
   \titlerunning{The peculiar galaxy NGC~128}
   \authorrunning{D'Onofrio \etal}
   \maketitle
 
   \begin{abstract} 

This is a multiband photometric and spectroscopic study of the peculiar 
S0 galaxy \object{NGC~128}. We present results from broad ($B$ and $R$) and
narrow (\ha) band optical CCD photometry, near (NIR) and far (FIR) infrared
observations, long slit spectroscopy, and Fabry-Perot interferometry (CIGALE).

The peculiar peanut shape morphology of the galaxy is observed both at optical
and near-infrared wavelengths. 
The stellar disk is thick and distorted (arc-bended), with a color
asymmetry along the major axis due to the presence of a large amount
of dust, estimated through NIR and FIR data of $\sim 6\times10^6 M_{\sun}$,
in the region of interaction with the companion galaxy \object{NGC~127}.

The color maps are nearly uniform over the whole galaxy, but for the major axis
asymmetry, and a small gradient toward the center indicating the presence of a
redder disk-like component.
The \ha\ image indeed reveals the existence of a tilted gaseous ``disk''
around the center, oriented with the major axis toward the companion galaxy
NGC~127. 

Long slit and CIGALE data confirm the presence of gas in a disk-like component
counter-rotating and inclined approximately of $50^\circ$ to the line of
sight. The mass of the gas disk in the inner region is $\sim 2.7\times10^4
M_{\sun}$.

The stellar velocity field is cylindrical up to the last measured points of
the derived rotation curves, while the velocity dispersion profiles are
typical for an S0 galaxy, but for an extended constant behaviour along the
minor axis.
 

   \end{abstract}

\section{Introduction\label{Intro}}
The connection between the boxy/peanut morphology of galactic bulges 
and the presence of a bar is difficult to prove observationally.
The main difficulty comes from the fact that the peculiar morphology can
be seen only in almost edge-on systems.

Photometric evidences of a bar in boxy/peanut galaxies have been found only in
a few cases, \eg\ the barred galaxy \object{NGC~4442} at intermediate
inclination (Bettoni \& Galletta \cite{bet:gal}), the edge-on S0 galaxy 
\object{NGC~1381} (de Carvalho \& da Costa \cite{decarv}), \object{NGC~5170}
(Dettmar \& Barteldrees \cite{dett:bart}), \object{NGC~2654}, and 
\object{NGC~4469} (Jarvis \cite{jarv1}).

The kinematic observations offer further evidences of this connection.
Kuijken \& Merrifield (\cite{kuij:merr}) obtained a characteristic
``figure-of-eight'' rotation curve, which is likely the strong signature for the
presence of a bar, for two peanut shape objects: \object{NGC~5746} and 
\object{NGC~5965}.
Other known cases of galaxies with these peculiar rotation curves are
\object{IC~5096} (Bureau \& Freeman \cite{bur:free}), \object{NGC~2683}
(Merrifield \cite{merrif}), \object{NGC~5907} (Miller \& Rubin
\cite{mill:rub}, and \object{UGC~10205} (Vega \etal\ \cite{vega}).

Of course, given the small number of cases, new accurate photometric and
spectroscopic observations are necessary for solving the peanut/bar connection.

Here we present a large data-set for the most representative object of this
class: NGC 128, the standard peanut-shaped galaxy (Sandage \cite{sandage}).

In the following sections we describe the peculiar morphology of NGC~128 and
its group, present a 2D approximated model of the surface brightness of the
galaxy (Sec. \ref{morphS0}), give the results of long slit and CIGALE data
(Sec. \ref{Spectr}), show the surface brightness and color profiles of the
galaxy at optical and near infrared wavelengths (Sec. \ref{Photom}), present
the results of \ha\ and NIR observations (Sec. \ref{halpha} and Sec.
\ref{NIR}), analyze the far infrared IRAS data (Sec. \ref{FIR}), and briefly
summarize the whole datasets (Sec. \ref{Discuss}). The technical solutions
adopted for the data reduction are given in the Appendix.

\section{Morphology\label{morphS0}}
The peculiar edge-on S0 galaxy NGC~128 is the dominant member of a group
which includes \object{NGC~126} (E/SB0), NGC~127 (Sa), \object{NGC~130}
(E5), a faint anonymous Sa galaxy $5\farcm6$ North and $9\farcm5$
preceding NGC~128, and possibly NGC~125 (S0 pec) (Zwicky \etal\ \cite{zwick}).
The membership of \object{NGC~125} is controversial since its recession
velocity is higher than average by $\sim1000$ \kms.

Burbidge \& Burbidge (\cite{burb:burb}), Hodge \& Merchant (\cite{hodge}), and
Bertola \& Capaccioli (\cite{bert:cap}, BC77) provided the first data for
NGC~128 and its group.  The galaxies of this group are late Ellipticals and
early Spirals, and the recession velocities are in the range 4200--4600 \kms.
NGC~128 is connected by a bridge to NGC~127, and according to BC77 a set of
filaments protrudes southwards to NGC~126.  The galaxy NGC~125 is surrounded
by a large ring (BC77 estimated a diameter of $\sim 70$ kpc) which is
asymmetrically placed with respect to the center of the galaxy.

NGC~128 is classified as BS-I in the list of de Souza \& dos Anjos
(\cite{desouza}). The peculiar peanut shape of the bulge is clearly visible
even in the innermost isophotes: the bulge appears squared by four symmetric
bumps, forming an {\rm X}-structure with arms at $\sim45^\circ$ from the major
axis of the galaxy and extended, in projection, for $\sim20''$ (5.4
kpc)\footnote{We assume throughout the paper a distance of 56 Mpc for the
galaxy and $H_0 = 75$ \kmsM.}. The peculiar morphology is not due to
extinction effects (see next sections).

As suggested by Pfenniger \& Friedli (\cite{pfen:fri}) the {\rm X}-structure
is likely an optical illusion. We clearly see this effect in the original
frame by changing the cuts:
a thin disk and a small, and approximately round, bulge are seen in the
center at the higher counts, while an increase of the
thickness of the disk is apparent at lower counts.  Such ``flaring''
seems to originate the {\rm X}-structure. The impression one has is
that of looking to a ``papillon''. The stars seem pulled out of the
disk, the maximum effect taking place at $\sim11$ arcsec, corresponding
to $\sim3$ kpc. 

The galaxy is seen approximately edge-on. We derived a $\log(a_{25}/b_{25})$
ratio of 0.76, a value which is in relative good agreement with that found by
Guthrie (1992) in his sample of edge-on galaxies for the S0 class.
The observed major axis diameter is $\sim 50$ kpc (at $\mu_B=26.0$ \mga), and
the thickness along the minor axis $\sim 21$ kpc.

The disk appears bended toward West on both sides, either in the visual and in
the NIR images, in particular toward the South-West direction, where the
bending starts at $r = 37''$ (10 kpc) from the center. The peaks of the
light distribution along cuts perpendicular to the disk major axis have a
maximum shift of $7''$ (1.9 kpc).

\begin{figure}
\resizebox{\hsize}{!}{\includegraphics{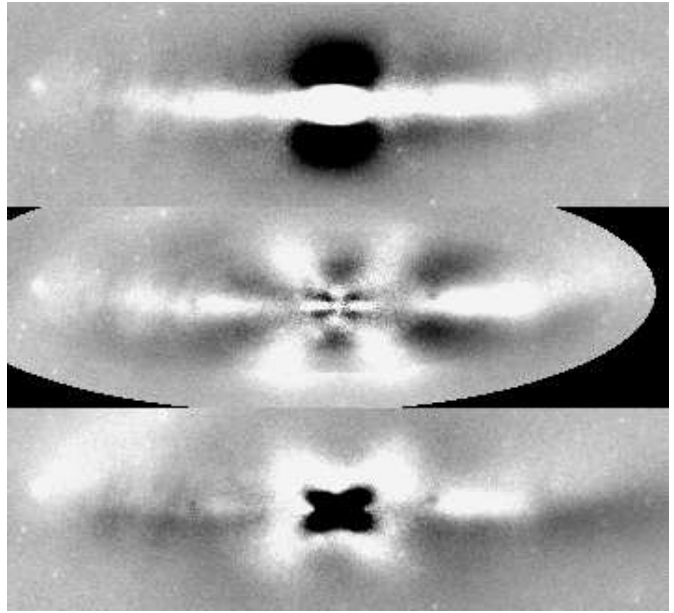}}
\caption{$B$-band maps of the O-C residuals for NGC~128:
{\it Upper panel}: the result of the unsharp masking technique;
{\it Middle panel}: the {\rm X}-structure and the disk of NGC~128 after
the subtraction of a model of the bulge component, obtained by fitting
the inner isophotes with ellipses; {\it Bottom panel}: the result of the
subtraction of a model of the bulge and disk component (see text). North is
on the left, West is up. The image is $\sim 100''$ in the N-S direction.}
\label{xstru}
\end{figure}

We tempted to highlight the peculiar morphology of NGC~128 in different
ways using the images with the best seeing and higher S/N ratio (the $B$-band
frames).
First we subtracted a model of the bulge component of the
galaxy (Fig.~\ref{xstru}, middle panel) obtained by fitting with
ellipses the inner isophotal contours of the galaxy.  In the residual
map we identify the {\rm X}-structure (in white color) and the disk
component, which appears thicker in the outer region.  The {\rm
X}-structure contributes approximately only to $\sim5\%$ of the total
luminosity of the galaxy. This means that it is a projection
effect rather than a physical new component.

While the {\rm X}-structure and the bending of the disk are clearly
visible in the original CCD frames, the increasing thickness of the
disk in the outer region is possibly the result of the bulge
subtraction (we verified this fact by simulating an edge-on S0 galaxy
and subtracting the bulge component).  Using the unsharp masking
technique (Fig.~\ref{xstru}, upper panel) we were able to confirm the
bending of the disk on both sides, but not the flaring of the outer
disk.  The {\rm X}-structure is obviously not visible in such image.

Notice that in Fig.~\ref{xstru} (middle and lower panels) the disk is fainter
along the North direction. Here there is a strong absorption in correspondence
of the encounter of the disk of NGC~128 with the arm of NGC~127.

Since the {\rm X}-structure is the result of the subtraction of
a model built with elliptical isophotes on a galaxy with a pronounced boxy
shape, we realized a new 2D model of the galaxy taking into account the
boxiness of the bulge.  The 2D surface brightness distribution of the entire
galaxy (within $\mu_B < 25$ \mga, masking the center, the
region of the disk affected by the interaction with NGC~127, and the distorted
outer disk) is given by the formulae:
\begin{eqnarray}
\mu_b &\!\!=\!\!& \mu_e + k \times [(R_b/R_e)^{1/n} -1] \ \ \ (bulge)\\
\mu_d &\!\!=\!\!& \mu_0 + 1.086 \times (R_d/h)          \ \ \  (disk)
\end{eqnarray}
where $R_b = \{[(x_i)]^c + [y_i/(b/a)_b]^c\}^{(1/c)}$ and
$R_d = \{[(x_i)]^2 + [y_i/(b/a)_d]^2\}^{(1/2)}$ are respectively the distance
of the pixel $(i,j)$ from the galaxy center. The exponent $c>2$ used for
the bulge component realizes the boxy isophotes.
The light profile of the bulge follows a $r^{1/n}$ law (Caon \etal\
\cite{caon2}), while the disk has an exponential distribution.

\begin{figure}
\resizebox{\hsize}{!}{\includegraphics{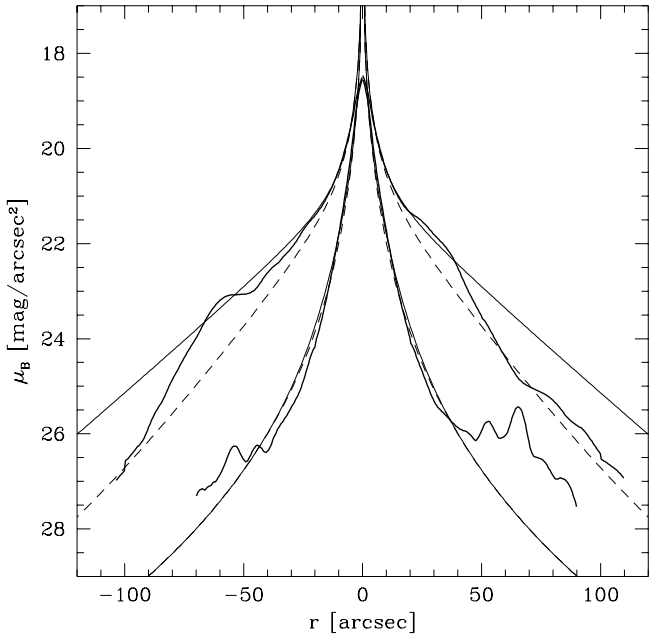}}
\caption{The 2D models of NGC~128 with superposed the major and minor axes 
$B$-band light profiles (thick solid lines). The thin solid line is our
$\chi^2$ solution, while the dashed line shows the fit realized by eye. North
and East directions are in the left part of the diagram}
\label{modfit}
\end{figure}

The model has the following best fitting structural parameters: $\mu_0 =
20.8\pm0.1$ \mga, $h = (24.5\pm0.5)''$, $\mu_e = 21.0\pm0.1$ \mga, $R_e =
(9.1\pm0.5)''$, $(b/a)_b = 0.67\pm0.05$, $(b/a)_d = 0.17\pm0.05$, $n =
4.7\pm0.5$, and $c = 3.69\pm0.05$.
This is, of course, only an approximated model for such peculiar galaxy
having a strongly disturbed morphology. 
The $\chi^2$ fit appears infact too bright in the center, even after 
the convolution with the PSF of the image (Fig.~\ref{modfit}) and, in 
the South direction, the true light distribution along the major axis is 
progressively fainter for $r>40''$, while along the minor axes 
the model is slightly brighter than the galaxy.  
Given the large uncertainty, we performed a second fit by eye, giving more
weight to the less disturbed Southern region of the galaxy. This provided a
smaller value for the scalelength $h$ of $17''$.

Even with the use of a more complex model, the residuals present an {\rm
X}-structure (Fig.~\ref{xstru}, lower panel), a disk obscured in the North
side, and a peculiar distorsion in the South-West direction.  The {\rm
X}-component now contributes only to $\sim 2\%$ of the galaxy luminosity.
This proves that the {\rm X}-structure is an optical illusion.

\begin{figure}
\resizebox{\hsize}{!}{\includegraphics{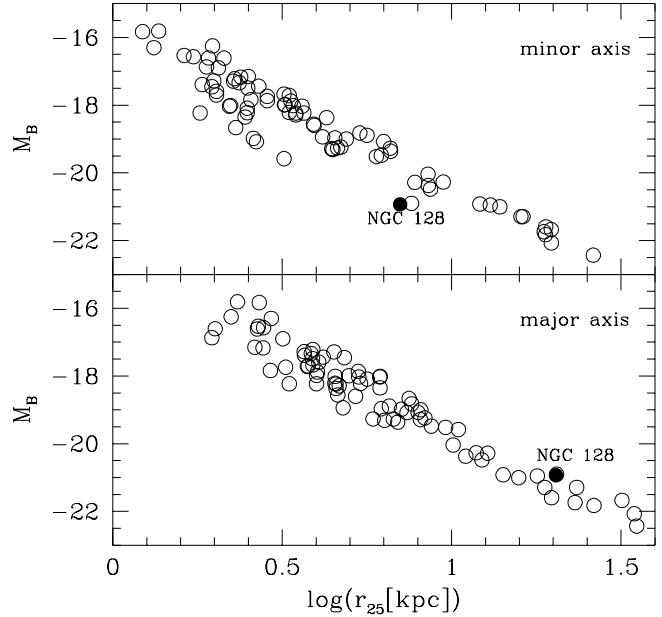}}
\caption{Position of NGC~128 in the $M_B-\log(a_{25})$ and
$M_B-\log(b_{25})$ diagrams (lower and upper panels respectively). In the same
plot we included for comparison the E and S0 galaxies of the Virgo cluster
from the C$^2$D sample}
\label{a25}
\end{figure}

\begin{figure}
\resizebox{\hsize}{!}{\includegraphics{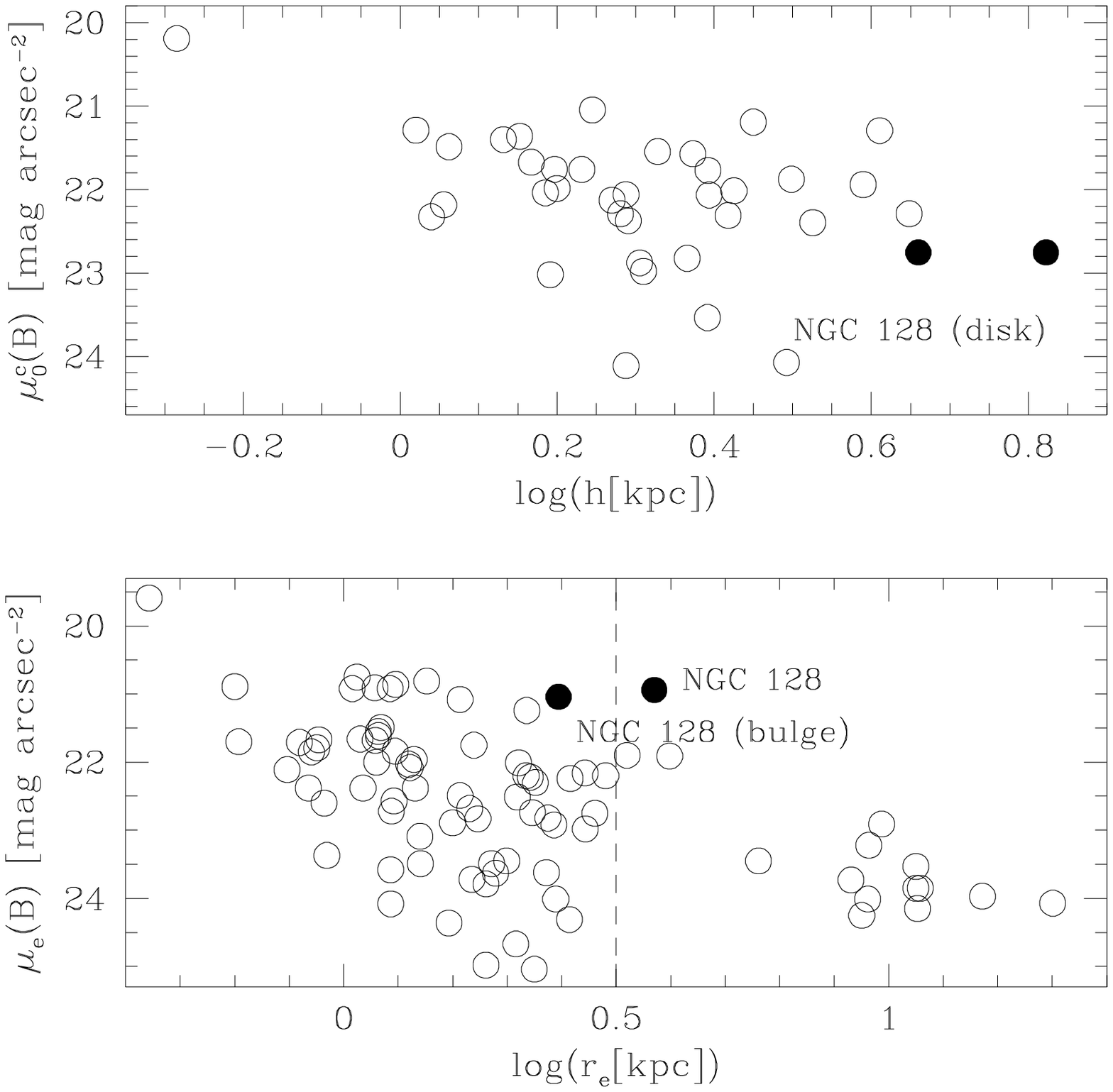}}
\caption{{\it Upper plot}: The central surface brightness (corrected for
inclination) and the scale-length of the disk of NGC~128 compared with a 
sample of 35 spiral galaxies of the Virgo cluster. The solid dots give the two
results obtained from the fit of the galaxy for the scalelength $h$. 
{\it Bottom plot}: Position of NGC~128 in the $\mu_e - \log(r_e)$ diagram.
The solid dots show the values obtained for the whole galaxy and for the
bulge component only. The error bars, non plotted here, are of the order of
the dot sizes.}
\label{muemu0}
\end{figure}

In Figs.~\ref{a25} and \ref{muemu0} we compare the 
structural parameters of NGC~128 with those extracted from a volume
limited sample of elliptical, S0, and spiral galaxies of the Virgo and
Fornax clusters (Caon \etal\ \cite{caon1}, \cite{caon3} (C$^2$D), D'Onofrio
\cite{donof1}). In terms of luminosity the galaxy belong to the `bright' family of
early-type objects defined by Capaccioli \etal\ (\cite{cap:etal}).
The major and minor axes, measured by the parameters $a_{25}$ and $b_{25}$,
are in good agreement with the corresponding data of the C$^2$D
sample for a galaxy of that luminosity (Fig.~\ref{a25}).
The effective surface brightness (and the effective radius) of the whole
galaxy and of its bulge component is relatively high: note in fact the
peculiar position in the $\mu_e - log(r_e)$ diagram (Fig.\ref{muemu0} lower
panel, \cf\ Capaccioli \etal\ \cite{cap:etal}) which would assign the object to 
the `ordinary' family of early-type objects.

For what concern the disk component, we measured the central surface
brightness and the scale length of the disk and plotted them in the $\mu^c_0 -
\log(h)$ diagram (Fig.\ref{muemu0} upper panel) comparing NGC~128 with a
sample of 35 spiral galaxies of various morphological types (D'Onofrio
\cite{donof1}). The scale length of the disk appears very large while the
corrected central surface brightness is normal for a galaxy of that luminosity.

The basic data for NGC~128 either derived in this work and extracted from
the literature have been listed in Table~\ref{ngc128}.
In the table $r_e$ and $r'_e$ are the radii of the circles which enclose half
the total luminosity of the galaxy. The first was calculated by excluding the
contribution of the two nearby companion galaxies of NGC~128. The HI mass
has been calculated by us following Jura (\cite{jura}).

\begin{table*}
\caption[]{The data for NGC~128.}
\begin{tabular}{llll}
\hline\hline\noalign{\smallskip}
Parameter &   & Value & Source \\
\noalign{\smallskip}\hline\noalign{\smallskip}
Coordinates (2000.0) & R.A. & 00$^h$ 29$^m$ 15$^s$.0 & de Vaucouleurs \etal\ 
(\cite{rc3}) \\
                     & Dec. & 02\degr\ 51$'$ 55$''$    & '' \\
Morphological type   & T    & $S0_1pec$                & Sandage (\cite{sandage}) \\
Heliocentric recession velocity & cz$_{hel}$ & 4180 $\pm$ 50 \kms\ & this paper \\ 
Distance             &$d=c z_{Hel}/H_0$& 56 Mpc                 & '' \\
Total $B$ mag.      & B$_T$& $12.87\pm0.1$ mag              & '' \\
Total $R$ mag.      & R$_T$& $11.23\pm0.1$ mag              & '' \\
$J$ mag. ($<96''$)  & J($96''$)&  $9.34\pm0.05$ mag              & '' \\
$H$ mag. ($<91''$)  & H($91''$)&  $8.51\pm0.05$ mag              & '' \\
$K$ mag. ($<90''$)  & K($90''$)&  $8.18\pm0.05$ mag              & '' \\
$(B-V)_T$ color     & $(B-V)_T$ & $1.02\pm0.01$ mag & de Vaucouleurs \etal\ 
(\cite{rc3}) \\
Average $(B-R)$ within \re\ & $\langle B-R \rangle$ & $1.68\pm 0.03$ mag & this paper \\
Average $(B-J)$     & $\langle B-J \rangle$ & $3.41\pm 0.06$ mag & '' \\
Average $(B-H)$     & $\langle B-H \rangle$ & $4.30\pm 0.05$ mag & '' \\
Average $(B-K)$     & $\langle B-K \rangle$ & $4.50\pm 0.07$ mag & '' \\
Galactic extintion     & $A_B$ & 0.06 mag  & de Vaucouleurs \etal\ (\cite{rc3}) \\
Effective radius & $r_e(B)$ &  $(13.9\pm0.5)''$  & this paper \\
                 & $r'_e(B)$ & $(19.2\pm0.5)''$  & '' \\
Effective surf. brightness & $\mue(B)$  & $20.94\pm0.1$ \mga\   & '' \\
                           & $\mue'(B)$ & $21.38\pm0.1$ \mga\   & '' \\
Radius at $\mu_B = 25$ \mga\ & r$_{25}$ North &  $(81.5\pm0.3)''$ & '' \\
                             & r$_{25}$ South &  $(69.3\pm0.3)''$ & '' \\
                             & r$_{25}$ East  &  $(25.7\pm0.3)''$ & '' \\
                             & r$_{25}$ West  &  $(27.4\pm0.3)''$ & '' \\
Major axis position angle ($\mu_B = 25$) &  P.A. &$0^\circ \pm1^\circ$   & '' \\
Central velocity dispersion & $\sigma_0$ & ($230\pm10$) \kms & '' \\
$V/\sigma$ ratio            & $V_{max}/\sigma_0$ & $1.52\pm0.12$ & '' \\
IRAS 12 \micron Flux & $F_{12}$ & upper limit & IPCS and FSC \\
IRAS 25 \micron Flux & $F_{25}$ & upper limit & IPCS and FSC \\
IRAS 60 \micron Flux & $F_{60}$ & $0.68\pm 0.07 Jy$ & FSC \\
IRAS 100 \micron Flux & $F_{100}$ & $1.69\pm 0.17 Jy$ & IPCS \\
\ha\ $+$ [NII] luminosity & $L(\ha + [NII])$ & $(2.18\pm0.41)\times10^{40}$ 
{\rm erg sec}$^{-1}$ & this paper \\
\ha $+$ [NII] mass & M(\ha) & $(2.7\pm1.3)\times10^4 M_{\sun}$ & '' \\
FIR luminosity & $L_{FIR}$ & $4.2\times10^9L_{\sun}$ & '' \\
Integ. CO intensity  & I(CO)  & $<2.32$ K \kms\ & Taniguchi \etal\ 
(\cite{tan:etal})\\
HI flux    & HI        & $< 1.4$ Jy \kms\ & Chamaraux \etal\ (\cite{cham:etal})\\
HI Mass    & M(HI) & $< 1.2\times10^9$ M$_{\sun}$ & '' \\
HI Mass    & M(HI) & $3.7\times10^9$ M$_{\sun}$ & this paper \\
X-ray luminosity  & $\log(L_X)$ & $< 41.51$ {\rm erg sec}$^{-1}$ & 
Fabbiano \etal\ 
(\cite{fab:etal})\\
Radio flux density & $L(3.6 cm)$ & $< 0.13$ mJy/beam & Collison \etal\ 
(\cite{coll:etal}) \\
\hline\hline\noalign{\smallskip}
\end{tabular}\label{ngc128}
\end{table*}

\section{Spectroscopic data\label{Spectr}}
\subsection{Stellar kinematics}
\begin{figure}
\resizebox{\hsize}{!}{\includegraphics{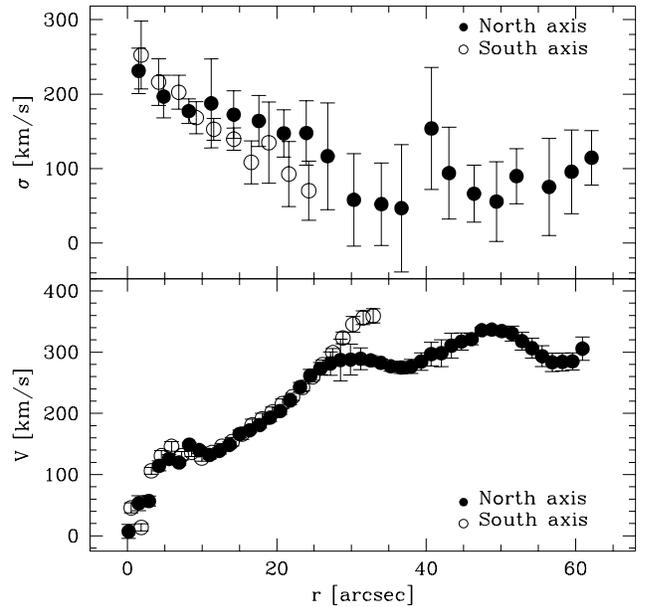}}
\caption{{\it Upper panel}: The folded velocity dispersion profiles along the
major axis of NGC~128. {\it Bottom panel}: The folded RCs along the major axis.
The South and North axes are indicated by open and filled circles
respectively}
\label{folded}
\end{figure}

Details of the available spectra and of the data reduction are given in 
\ref{SPM}. Here we discuss the rotation curves (RCs) of the star
component extracted from the absorption features of the spectra.
The radial velocity measurements of the spectra along the major
axis were combined and folded about the center of the galaxy and shown in
Fig.~\ref{folded} (bottom panel), choosing 
the radial offset and the systemic velocity which minimize the dispersion
of points in the folded curves ($0\farcs6$ along the South direction
corresponding to $\sim150$ pc). 
Our heliocentric systemic velocity is $4180\pm50$ \kms,
lower by 30 \kms\ than in Dressler \& Sandage (\cite{dres:sand}).
The error takes into account the zero point uncertainty of the RC derived
from the three standards stars used.

Note that the folded curves match very well in the inner $\sim20''$, while
the curve along the North side suffer from the presence of the interaction
with NGC~127. The South side has a short extension because the RC
is extracted from the off-centered spectrum of the first night. The
spectrum of the second night (much more noisy) does not extend far out
Southwards. 

The RC is very steep in the central region: at a distance
of $5''$ from the center along the major axis the rotation velocity is
already 100 \kms. 
In BC77 the RC grows more slowly in the range $20'' \div 40''$, while in our
measurements the velocity increases outside $50''$ up to a value of $\sim$350
\kms\ in the North direction.

In NGC~128 we do not see in the RC the characteristic ``figure-of-eight''
feature, neither for the stellar or the gaseous component.
Here the gas is counter-rotating and we have an {\rm X}-shaped RC 
(see \ref{gaskin}). According to Friedli \& Udry (\cite{fri:udr}) and
Emsellem \& Arsenault (\cite{ems:ars}) the counter-rotating gas is tracing
the anomaouls orbits existent in a tumbling triaxial potential.

\begin{figure}
\resizebox{\hsize}{!}{\includegraphics{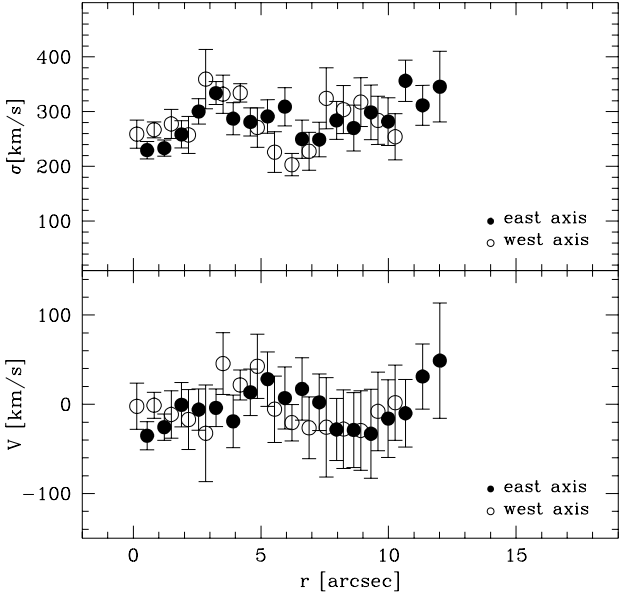}}
\caption{{\it Upper panel}: The folded minor axis velocity dispersion
profiles of NGC~128. {\it Bottom panel}: The folded minor axis RCs. Open and
filled circles indicate the East and West side respectively}
\label{minax}
\end{figure}

The folded RC along the minor axis of the galaxy is shown in
the bottom panel of Fig.~\ref{minax}. 
The curve along the minor axis is less extended since
beyond $\sim15''$ the spectra reach the level of the sky surface brightness
and the errors become larger.
There is a hint of a non-zero velocity pattern along the minor axis
suggested by the occurrence of a maximum and of a minimum velocity at
symmetric places along the two sides opposite to the center. 
If this behaviour will be confirmed by future data, 
the presence of a small ring of stars (remnants of a polar ring?) can be
suspected.

\begin{figure}
\resizebox{\hsize}{!}{\includegraphics{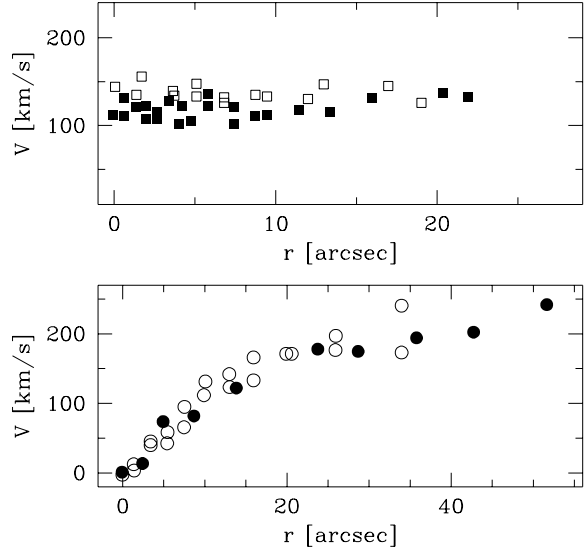}}
\caption{{\it Upper panel}: The RCs of NGC~128 along cuts perpendicular
to the major axis. Filled square mark the RC at $r=10''$. Open square at
$r=15''$. {\it Bottom panel}: The RCs along cuts parallel to the major axis:
filled circles ($r=4''$), open circles ($r=8''$)}
\label{jarvisb}
\end{figure}

A number of off-centered spectra of NGC~128, along directions
parallel and orthogonal to the main axes, have been obtained by B. Jarvis
(private communication). We list the Jarvis' logbook in Table~\ref{BJarvis}
and we plot the corresponding RCs in Fig.~\ref{jarvisb}.
The agreement with our data is quite good. 
The comparison of the measured velocities at a given distance $r$ along the RC
is shown in Table~\ref{BJ}.

\begin{table}
\caption[]{B. Jarvis long-slit observations of NGC~128.}
\begin{tabular}{lllllr}
\hline\hline\noalign{\smallskip}
Telescope & Instr. & Disp.& PA & Exp.T.& Notes \\
          &        & Ang/pxl & [deg] & [hour] & \\
\noalign{\smallskip}\hline\noalign{\smallskip}
CTIO 4m & SIT       & 0.94 & 90 & 1.5 & $15'' ~~ \perp$ S \\
CTIO 4m & 2DF       & 0.55 & 90 & 2.7 & $10'' ~~ \perp$ S \\
CTIO 4m & 2DF       & 0.55 &  0 & 3.3 & $8'' ~~ \parallel$ E \\
AAT 3.9m& IPCS      & 0.49 &  0 & 1.6 & $4'' ~~ \parallel$ E \\
\hline\hline\noalign{\smallskip}
\end{tabular}\label{BJarvis}
\end{table}

Such spectra show that the cylindrical rotation
is observed up to $20''$ ($\sim 5.4$ kpc). The major axis off-set RCs
are also quite similar to our curve, differing for a smaller gradient only.   

\begin{table}
\caption[]{Comparison of our velocities with the data extracted from the RC of 
B. Jarvis.}
\begin{tabular}{l|ll}
\hline\hline\noalign{\smallskip}
$r$ & $V_{our}$ (\kms) & $V_{BJ}$ (\kms) \\
\noalign{\smallskip}\hline\noalign{\smallskip}
$10''$ & 138 & 120 \\
$15''$ & 142 & 140 \\
\hline\hline\noalign{\smallskip}
\end{tabular}\label{BJ}
\end{table}

The upper panel of Fig.~\ref{minax} shows the folded minor axis velocity
dispersion (VD) profile.
The central value is around 220--240 \kms. 
Note the increase of $\sim100$ \kms\ in the inner $3''$ and the wave-shape
which keeps the velocity dispersion to an high level out to the last
measurable point.

The folded VD profile along the major axis is shown in
Fig.~\ref{folded} (upper panel). The shape is that characteristic of the
early-type galaxies, with a bulge dominated region where the velocity
dispersion decreases, and a disk dominated part, where the velocity remains
appromimately constant.
It is interesting to note the asymmetry in the VD profile
at $\sim40''$ in correspondence of the arm of NGC~127.

\subsection{Gas kinematics}\label{gaskin}
Here we discuss the RCs of the gas component detected in our spectra
and the results of CIGALE observations (see \ref{SPM} and 
\ref{CIG} for details of the data acquisition and reduction).

\begin{figure}
\resizebox{\hsize}{!}{\includegraphics{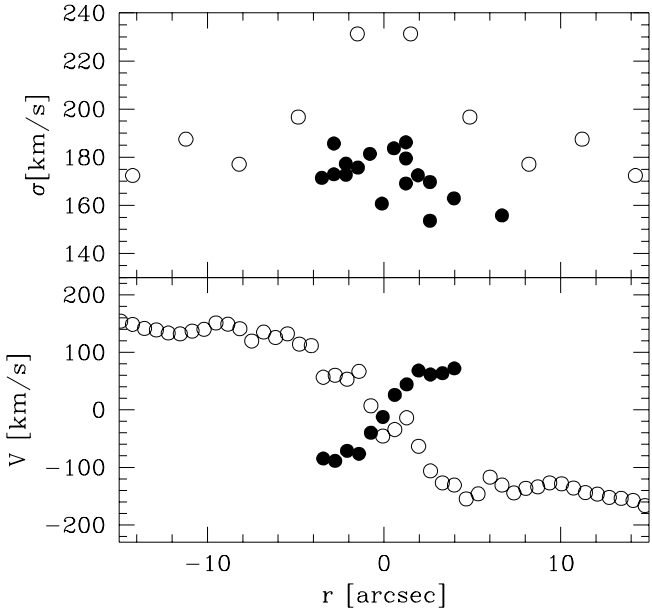}}
\caption{{\it Upper panel}: The gas velocity dispersion along the major axis 
of NGC~128 (filled circles) compared to the inner star velocity dispersion
(open circles) averaged over the two semiaxes. 
The error bars, non plotted here, are of $\sim 20$ \kms.
{\it Lower panel}: The counter-rotation of the gas component in NGC~128 along
the major axis. The long slit RC of the gas, derived from emission lines, is
marked by filled circles. The RC of the stars is plotted with open circles.
North is on the left. The error bars, not plotted here, are
comparable to the dot sizes}
\label{rcgas}
\end{figure}

In Fig.~\ref{rcgas} together with the unfolded RC obtained
from the absorption lines, we plotted the behaviour of the gas component 
resulting from the emission lines.
We took the peak of a gaussian curve, fitted to the emission lines in each row
of the spectra, as a measure of the rotational velocity of the gas.
It appears a clear counter-rotating gas component which extends up to
$\sim8''$ (2.2 kpc) around the nucleus. 
The gas seems to have the same gradient of the stellar component. In the first
$4''$ around the nucleus the rotation velocity increases up to $\sim 
100\div120$ \kms. This behaviour is in agreement with the velocity field
derived by Emsellem \& Arsenault (\cite{ems:ars}) with TIGER. They found that
the gas and the stellar velocity at $3.5''$ along the major axis is $\sim 140$
\kms.

Unfortunately the S/N ratio is not high enough to follow the gas emission
at larger distances. We also do not observe the ``figure-of-eight''
in the rotation curve which is a strong signature of a barred potential
(Kuijken \& Merrifield \cite{kuij:merr}).

The velocity dispersion of the gas is more difficult to evaluate.
We derived an approximate value by correcting the sigma of the gaussian, used
to fit the emission lines, for the instrumental dispersion through the relation:
$\sigma_c = \sqrt(\sigma^2 - \sigma^2_{instr})$. The velocity dispersion
is nearly constant at $\sim 175$ \kms\ within the central $5''$.
This is only $\sim 55$ \kms\ lower than the central stellar velocity dispersion.
A possible explanation for this high value is that the gas is not 
in equilibrium yet.

\begin{figure}
\resizebox{\hsize}{!}{\includegraphics{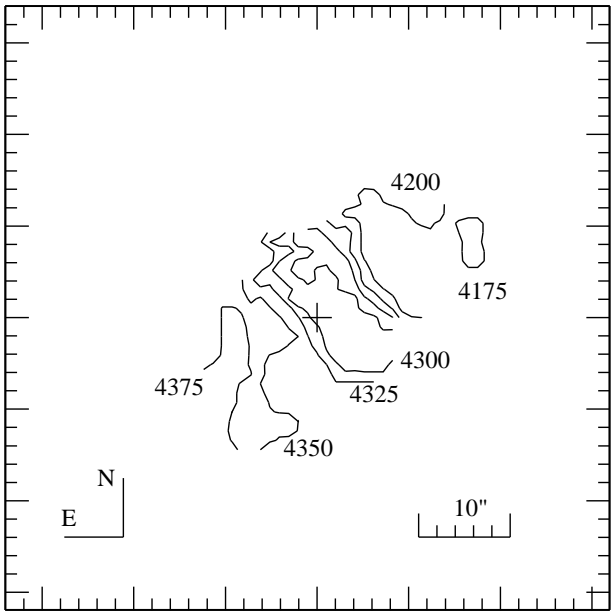}}
\caption{The velocity field of NGC~128 resulting from CIGALE. The center
is marked by a cross}
\label{rcfield}
\end{figure}

The velocity field of NGC~128 derived by the CIGALE data is plotted
in Fig.\ref{rcfield}. It is well consistent with a disk-like gas 
component. The rotational velocity is positive along the SE direction and 
negative in the NW.
The major axis of the \ha\ disk is observed to extend up to $\sim25''$ and is
approximately oriented at a position angle PA $\sim 120^\circ$. The PA
decreases from the center (PA $\sim 130^\circ$) to the outer parts (PA $\sim
100^\circ$).

CIGALE is in poor agreement with the long slit spectroscopic data
(Fig.\ref{rcgas}). The gradient of the RC, the maximum rotational velocity,
and the systemic velocity of the galaxy (greater by $\sim 90$ \kms) do not
match the EFOSC data. The discrepancy is probably due to the loss of
resolution caused by the binning of the CIGALE data. On the other hand the
agreement is fair with the photometric observations, despite the lower
resolution and the bad seeing condition. The extension and the PA of the gas
disk component are similar.

From the 2D velocity field of the gas, following Plana \& Boulesteix 
(\cite{plana:boul}), 
we derived an inclination for the disk
of $\sim50^\circ \pm5^\circ$ which is in fair agreement with the value of
$\sim 60^\circ$ computed from the apparent flattening of the \ha\ image (see
\ref{halpha}).


\section{Optical photometry\label{Photom}}
Now we present the results of our $B$ and $R$ bands CCD photometry. Details
of the available data and the reduction procedures are given in \ref{Phot}.

Our CCD photometry is slightly fainter ($\sim 0.28$ mag) than that of BC77 who
claimed a corrected $B$-magnitude inside a $114''$ circular aperture of 12.61
mag, while it is 0.02 brighter than the value of 13.01 mag reported by de
Vaucouleurs \& de Vaucouleurs (\cite{rc2}).  By excluding the contribution of
the two nearby galaxies NGC~126 and NGC~127 and integrating the total
luminosity up to $169''$ we get $B=12.87\pm0.1$ mag (in good agreement with
the value of $12.77\pm0.13$ of de Vaucouleurs \etal\ \cite{rc3}) and
$R=11.23\pm0.1$ mag.  We obtained (with our choice of $H_0$) an absolute
magnitude $M_B \sim -20.93$.  Most of the results of our photometric analysis,
together with a number of literature data, are summarized in
Table~\ref{ngc128}.

The isophotes of the galaxy are nearly round in the center
and become progressively influenced by the presence of the disk after few 
arcsec. The simmetry of the figure is good and there are no indications of
subcomponents and/or substructures.
A first set of profiles was extracted (in each photometric band) along the 
major, minor and intermediate axes of NGC~128. 
We averaged 10 profiles taken in a cone of 1 degree aperture. 

\begin{figure}
\resizebox{\hsize}{!}{\includegraphics{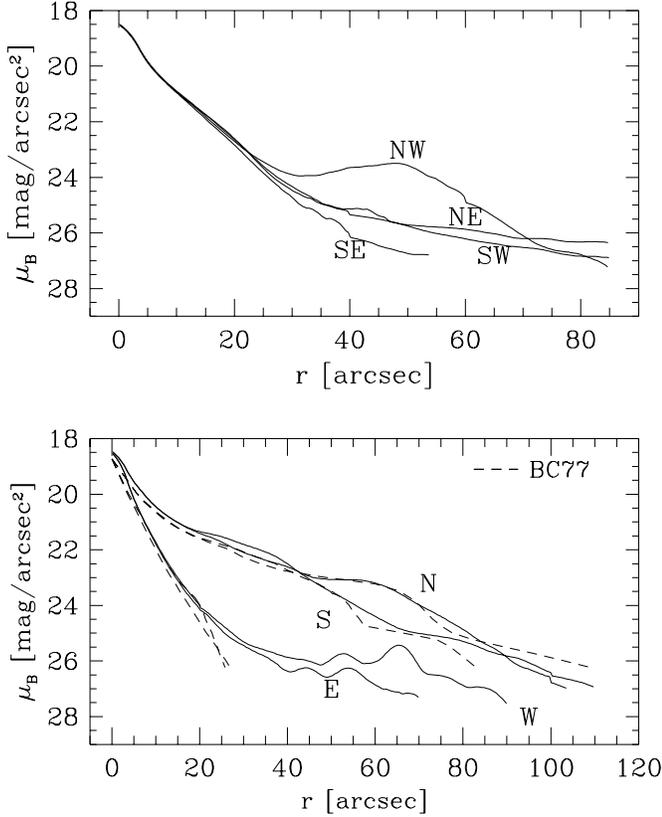}}
\caption{{\it Upper panel}: The intermediate axes profiles of NGC~128.
{\it Lower panel}: Observed major and minor axes profiles of NGC~128.
The 1$\sigma$ error is $\sim0.02$ \mga\ at $\mu_B = 18$ and increases up to
$\sim0.25$ \mga\ at $\mu_B = 26$. 
The dashed lines represent the profiles obtained by BC77}
\label{majminB}
\end{figure}

The $B$-band folded light profiles of NGC~128 are plotted in Fig.\ref{majminB}. 
The asymmetry along the major axis starts approximately at $20''$ and it 
is clearly seen between $50''$ and $90''$. On the other hand, along the
minor and intermediate axes the light profiles are more symmetric 
and become progressively asymmetric because of the presence of
the nearby galaxies NGC~126 and NGC~127. 

The profiles given by BC77 are shown in the figure with the dashed lines. A
small systematic offset seems to exist between the two data sets, but on
average the agreement is very good, taking into account that they used
photographic data.

\begin{figure}
\resizebox{\hsize}{!}{\includegraphics{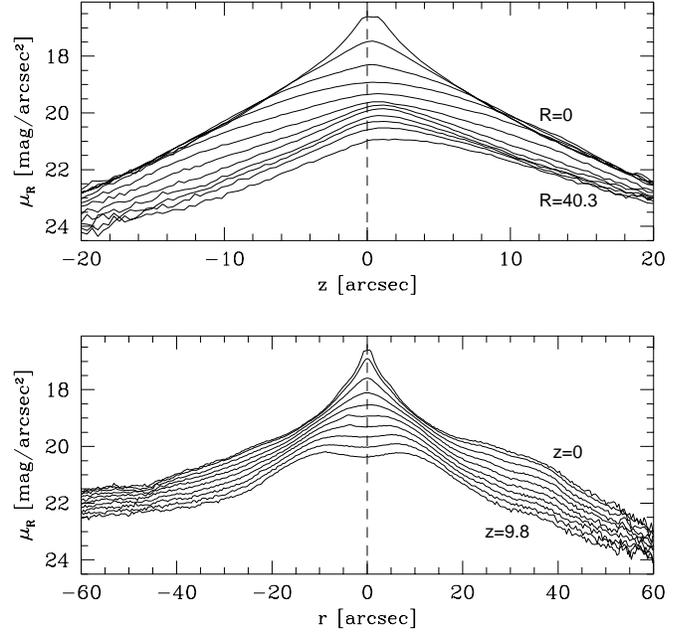}}
\caption{{\it Upper panel}: The light profiles of NGC~128 along cuts 
perpendicular to the major axis (South direction). West is on the right
side. {\it Bottom panel}: Profiles parallel to the major axis (East direction).
South is on the right side}
\label{rzp}
\end{figure}

A second set of profiles were extracted (for the $R$-band image) along cuts
perpendicular and parallel to the major axis in the East and South directions
respectively. Note the shift of the peak of surface brightness toward West
(Fig.~\ref{rzp} upper panel). From the bottom panel of that figure we have a 
marginal indication for the presence of a bar. Following Wakamatsu \& 
Hamabe (\cite{waka:ham}) (who presented the evidence for the presence of a bar
in the edge-on galaxy \object{NGC~4762}) we tentatively recognized in the
light profiles (along the South direction which is less affected by dust) the
typical bump usually attributed to the existence of a bar.
This is indicated by the fact that the bump becomes more and more insignificant
at larger $z$. Differently from NGC~4762 this behaviour is not followed by a
plateau and a cut-off characteristics of the presence of a lens.
So, our suspect for the presence of a bar rests only on the marginal evidence
given in Fig.~\ref{rzp}.

\begin{figure*}
\resizebox{\hsize}{!}{\includegraphics{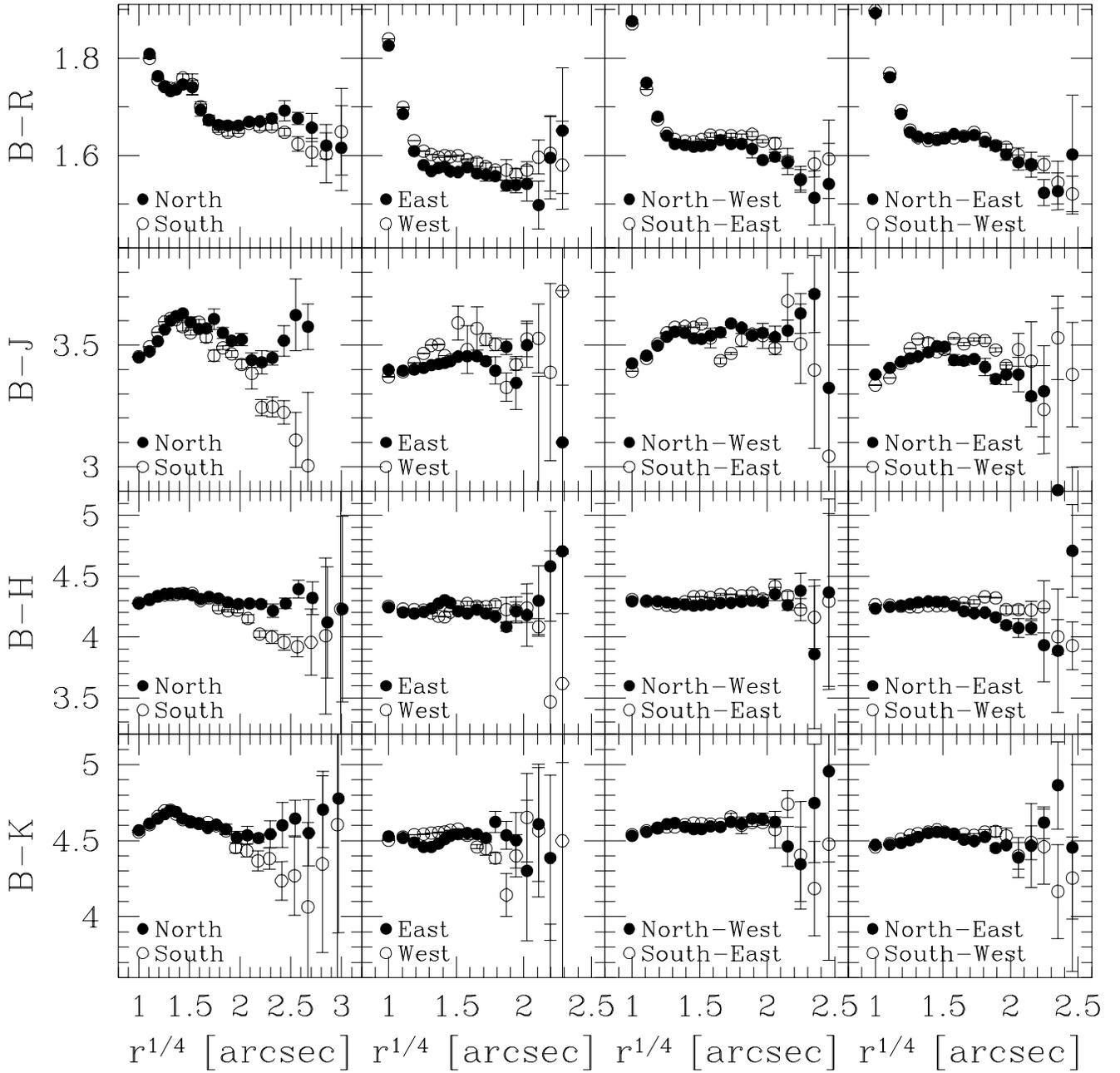}}
\caption{The $B-R$, $B-J$, $B-H$, and $B-K$ color profiles of NGC~128.
From left to right: the major axis, the minor axis, and the intermediate axes}
\label{majcol}
\end{figure*}

The $B-R$ folded color profiles of the major axis are shown in Fig.~\ref{majcol}
(together with the $B-J$, $B-H$, and $B-K$ colors). 
They have been preferred on a 2D color map in order to increase the S/N ratio.

We notice a clear reddening toward the center ($\sim 0.15$ mag)
starting approximately at $r\sim 10''$ and we estimate an average
color $\langle B-R \rangle = 1.68\pm0.03$. Such reddening is not symmetric
along the various direction of profile extraction, a fact that further support
the existence of a disk-like structure around the center.

The color profiles become bluer where the arm of NGC~127 encounters the disk of
NGC~128, the maximum shift between the two sides reaching 0.06 mag.
The $B-R$ color increases of $\sim0.20$ mag in the inner 15 Kpc. 
Along the minor and intermediate axes the $B-R$ color gradient is even
larger, reaching 0.3 mag in the direction of the {\rm X}-structure.

\section{Narrow band photometry\label{halpha}}
As before the details of the data reduction are given in 
\ref{alphaphot}.

NGC~128 is already known to possess a counter-rotating gas component (Pagan
\cite{pagan}, Emsellem \& Arsenault \cite{ems:ars}).  In Fig.~\ref{rcgas}
together with the unfolded rotation curve of the galaxy we plotted the
behaviour of the gas component resulting from the long slit data.  It appears
a clear counter-rotating gas component which extends up to $r\magcir8''$ (2.2
kpc) around the nucleus.

\begin{figure}
\resizebox{\hsize}{!}{\includegraphics{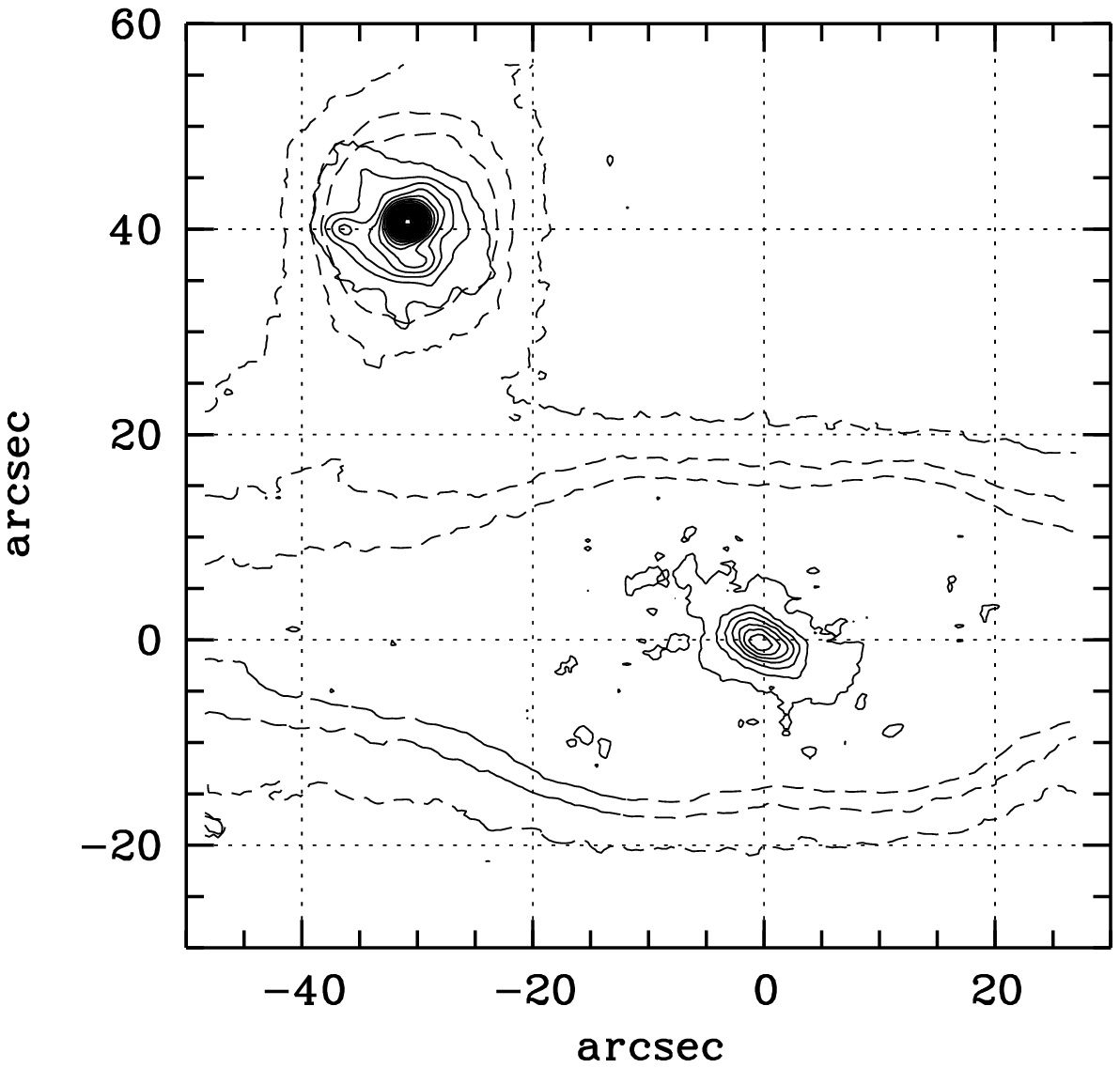}}
\caption{\ha\ and [NII] emission contours. The solid line gives the emission
coming from NGC~128 and NGC~127. The dashed line shows the contours of the
$B$-band image of the galaxies. North is on the left, West is up}
\label{halphaf}
\end{figure}

The contours of the \ha\ and [NII] emission lines of NGC~128 and
NGC~127 are shown as solid lines in Fig.~\ref{halphaf}. In the same figure
we overlaid the contours of the $B$-band image of the galaxy (dashed lines)
and a grid to help the evaluation of the size of the emission region. 

Notice the presence of an inner disk-like component with the major axis tilted
toward NGC~127. The disk is approximately at $\sim$40\degr\ from the equatorial
plane of NGC~128 and it is extended to $\sim10''$ from the center in both
directions.
The position angle of the gas component decreases toward the galactic plane
to $\sim$25\degr\ (in good agreement with the value quoted by Emsellem \& 
Arsenault \cite{ems:ars}).

The total \ha\ flux (computed in Appendix) was used to calculate the total mass
of the gas disk. Following Osterbrock (\cite{osterb}), Kim (\cite{kim})
derived the mass from the relation:
\begin{equation}
M_{HII} = 2.8\times10^2 \cdot \left(\frac{D}{10}\right)^2 
\cdot \left(\frac{F(\ha)}{10^{-14}}\right)
\cdot \left(\frac{10^3}{n_e}\right)
\end{equation}
where $M_{HII}$ is expressed in solar masses, $D$ is the distance of the galaxy
in Mpc, and $n_e$ is the electron number density in cm$^{-3}$, and F(\ha) is
given in {\rm erg sec}$^{-1}$ {\rm cm}$^2$.
Our final value is $M_{HII} = (2.7\pm1.3) \times10^4 M_{\sun}$, which rests on 
our value of $H_0$, an uncertainty of 0.4 Mpc for the distance of the galaxy,
on the measured flux error, and an assumed electronic density of $(10\pm
3)\times10^2 e^- cm^{-3}$.
This density was estimated following Kim (\cite{kim}) from the
ratio of the two lines $\lambda$ 6717/6731 ($\sim$ 0.75 a value which is
compatible with such electronic density) between the [SII] emission lines
detected in the spectra.

We also computed the gas mass using CIGALE observations. By integrating the
total $\ha\ + [NII]$ flux within $25''$ and assuming $n_e = 500 e^- cm^{-3}$
we get $M_{HII} = 7\times10^4 M_{\sun}$. Although the flux is integrated here
on a larger area, we believe it is likely an overestimated value,
due to the poor photometric calibration of the CIGALE data.

\section{Results from NIR data\label{NIR}}
As for the optical images, the NIR frames also show that the inner region have
the characteristic {\rm X}-structure and that the disk is arc-bended toward
West. Details of the data reduction are given in the \ref{Nirphot}.

We have extracted the light profiles along the same directions of the optical
images and built the $B-J$, $B-H$, and $B-K$ color profiles. This was realized
by degrading the $B$-band image to the same seeing of the NIR images.  They
are shown in Figs.~\ref{majcol}. The $(J-H)$, $(J-K)$, and $(H-K)$ colors (
non plotted here) are constant, within the present uncertainty, up to the last
measured point ($J-H = 0.70\pm0.05$, $J-K = 1.00\pm0.05$, $H-K = 0.30\pm0.05$).

A few comments are in order:\\
1) along the major axis the folded profiles do not match: the difference
reaching up to $\sim0.5$ mag in $(B-J)$, $(B-H)$, and $(B-K)$ at
$r\sim40''$ from the center;\\
2) in the inner $3''$ the $(B-J)$, $(B-H)$, and $(B-K)$ color profiles
along the major and intermediate axes have a small decrease ($\sim0.15$ mag),
which is less evident along the minor axis. This reveals the presence of an
elongated red component. In the same region the $(B-R)$ color increases;\\
3) apart from the $(B-R)$ color, the other colors are approximately constant
with the distance from the center.

From the color indices we derived information about the dust
distribution.  By comparing the North and South semiaxes, we
found that the color indexes $B-\lambda$ ($\lambda = V, R, J, H, K$) are
systematically higher in the North direction, between $30''$ and $60''$, than
in the South.
We interpret the asymmetry as due to an enhancement of the average dust
content in the region where the arm of the companion galaxy NGC~127 intercepts
the major axis of NGC~128.

We computed the color excess $E(B-\lambda)$ by subtracting the
corresponding color index of the two semiaxes, and, following the procedure 
described in \ref{ODM}, we derived a dust mass excess of 
$\sim6\times10^6$ M$_{\sun}$ associated to the North semiaxis.  
The IRAS data confirm that the dust is manly concentrated toward the
region of interaction of the two galaxies (see Sect. \ref{FIR}) and,
therefore, we may consider the derived dust mass as a measure of the dust
content of NGC~128.
Of course, adopting different models lower dust 
masses (up to one order of magnitude) can be obtained.

Taking into account the uncertainty in the color excess, the evaluation of
the dust mass from the optical and NIR data is characterized by an error of
$\sim40\%$, a factor of two larger than those characterizing the FIR 
measurements.

\section{The FIR emission\label{FIR}}
NGC~128 is associated in both the IRAS Point Source Catalogue (IPSC) and 
Faint Source Catalogue (FSC) to the IRAS source 00266+0235. At 12 and 
25 \micron\ the IRAS catalog gives only upper limits, while high quality
fluxes are available at 60 and 100 \micron\ respectively in the FSC and IPSC
(see Table~\ref{ngc128}). 
The total FIR luminosity, following Helou 
\etal\ (\cite{helou}) and adopting our heliocentric recession velocity and Hubble 
constant, turns out of $4.2\times10^9L_{\sun}$.

Knapp \etal\ (\cite{knapp:etal}) derived the fluxes at 60 and 100 \micron\ for
NGC~128 by averaging the IRAS data at the galaxy position. Using a similar
technique, Bally \& Thronson (\cite{bal:thr}) obtained flux values which are
slightly different. Their results are in agreement, within the errors, with
the PSC values reported in Table~\ref{ngc128}.

\begin{figure}
\resizebox{\hsize}{!}{\includegraphics{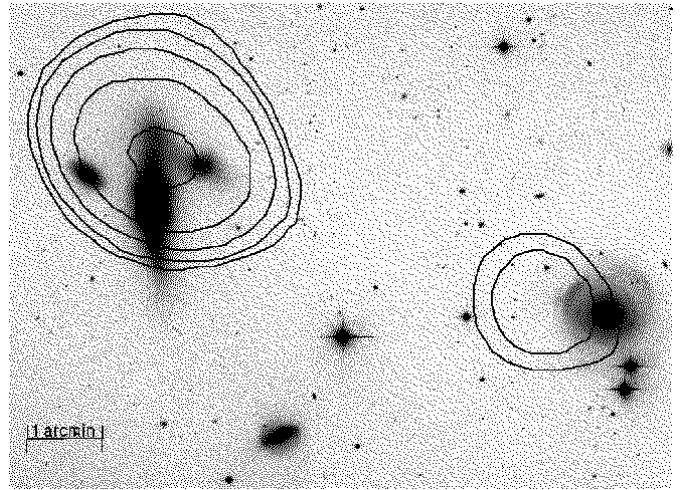}}
\caption{The group of NGC~128 with superposed the contours of the 
emission at 60 \micron\ detected by IRAS. North is up and East to the left}
\label{infrared60}
\end{figure}

\begin{figure}
\resizebox{\hsize}{!}{\includegraphics{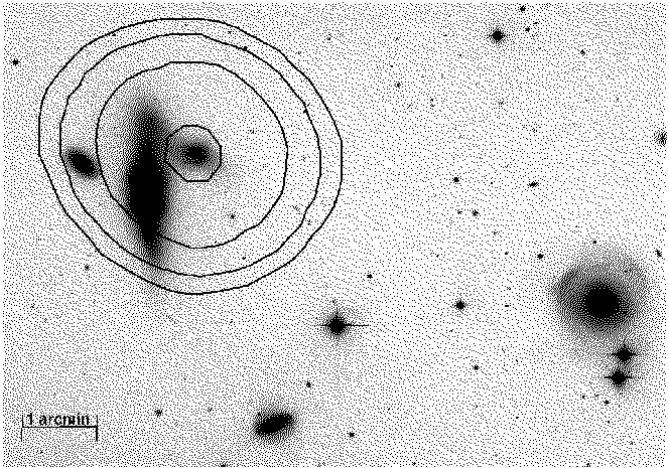}}
\caption{The group of NGC~128 with superposed the contours of the emission 
at 100 \micron\ detected by IRAS. North is up and East to the left}
\label{infrared100}
\end{figure}

The 60 and 100 \micron\ IRAS data were processed using the method described by
Assendorp \etal\ (\cite{ass:etal}). After the standard co-addition of the
images, we used the Maximum Entropy (ME) method described by Bontekoe \etal\
(\cite{bont:etal}). The resulting 60 and 100 \micron\ high-resolution (about
$1'$) infrared maps are overlaid onto a deep $V-$band image of NGC~128, taken
with CAFOS at the 2.2m telescope of Calar Alto (Figs.\ref{infrared60} and
\ref{infrared100}).

The peak of the FIR radiation is situated in the area where the galaxy
undergoes interaction with NGC~127.  The shift between the 60 and 100
\micron\ peak intensity position is $42''$ in the IRAS cross scan (SW)
direction and therefore not significant compared to the $1'$
resolution achieved with the ME method.  Taking into account this
uncertainty, we cannot associate the FIR emission to NGC~128 instead
of NGC~127. Both these galaxies may contribute to the FIR emission.

On the other hand, the shift in the NW scan direction is 
much smaller ($\sim 5''$) implying that the FIR radiation has its peak in the
North semiaxes, so that it is probably connected to the interaction of 
the two galaxies. 

We note in passing that the companion galaxy NGC~125 is not seen at 100
\micron, and that the ring-like structure visible at 60 \micron\ is
approximately in the same position of the optical ring detected by
BC77. The presence of such peculiar feature leads to suspect that
NGC~128 and NGC~125 possibly underwent an interaction in the past.

The FIR thermal emission can be used to measure the dust mass. 
The resulting dust mass depends on the physical-chemical properties of the
grains (\ie\ grain radius, density and emissivity) and 
on the adopted dust temperature. 

In principle we can derive the dust temperature ($T_d$) by
fitting the FIR data with a grey-body characterized by a spectral trend
$\lambda^{-\alpha}$. Since only the 60 and 100 \micron\ fluxes are available,
we used the relation obtained by Henning \etal\ (\cite{henn:etal}) to compute 
three dust color temperatures, obtaining respectively $T_{bb}=39$ \Kelv 
($\alpha=0$ black-body),
$T_{gb}=33$ \Kelv ($\alpha=1$ grey-body),
$T_{gb}=28$ \Kelv ($\alpha=2$ grey-body).
The values chosen for $\alpha$ 
are suggested by the current dust models (Mathis \&
Whiffen \cite{math:whi}; Wright \cite{wright}; D\'esert \etal\ \cite{desert};
Draine \& Malhotra \cite{drai:mal}).

Following Hildebrand (\cite{hild}) we computed the dust mass adopting a
spectral index $\alpha=1$ (except with the formula of Thuan \& Sauvage
\cite{thua:sau} where we used $\alpha=1.5$). The derived masses are listed in
Table~\ref{tabledust}. Notice that, considering the flux uncertainty of the
IRAS data ($\sim 10\%$), our values are in good agreement with those available
in literature. The differences with Bally \& Thronson (\cite{bal:thr}) and
Roberts \etal\ (\cite{rob:etal}) are due both to differences in the adopted
fluxes and in the dust temperature evaluations.

\begin{table}
\caption[]{Dust masses with a single temperature model.}
\begin{tabular}{cll}
\hline\hline\noalign{\smallskip}
 Dust mass& Method & Reference \\
($M_{\sun}$)& & \\
\hline\hline
&&\\
$1.8\times10^6$ & Greenhause \etal\ \cite{gree:etal} & Bally $\&$ Th. 1989 \\ 
&&\\
 $2.8\times10^6$ & Young \etal\ \cite{young} & Roberts \etal\ \cite{rob:etal} \\
&&\\
 $1.5\times10^6$ & Thronson $\&$ Telesco \cite{thr:tel} & this paper \\
& Greenhause \etal\ \cite{gree:etal} & \\
& Young \etal\ \cite{young} \\
&&\\
 $1.1\times10^6$ & Roberts \etal\ \cite{rob:etal} & this paper \\
&&\\
 $4.3\times10^5$ & Thuan $\&$ Sauvage \cite{thua:sau} & this paper \\
&&\\
\hline
\hline
\end{tabular}\label{tabledust}
\end{table}

The dust masses in Table~\ref{tabledust} are derived using a single
temperature model.  This is a rough approximation for the condition found
in the galactic environment.  The dust is in fact heated by the radiation
field, which in turn depends on the sources of luminosity and their spatial
distribution in the galaxy.  The total FIR emission is likely due to the
contribution of dust at different temperatures.  Moreover, the IRAS FIR
measurements are not adequate to detect the emission coming from cold dust
($10\div20$ \Kelv) which peaks at a wavelength between 200 and 300 \micron.

We thus estimate the total dust mass by introducing a dust temperature
distribution depending on two free parameters which affect the shape
of the function (see Kwan \& Xie \cite{kwan:xie} and Merluzzi \cite{merluz} 
for details).

We choose a range of temperatures that contribute to the IRAS emission
between 7 and 60 \Kelv\ and we select the proper values of the free parameters
which reproduce the ratio of the flux densities at 60 and 100 \micron\ and
takes into account the value of the FIR color temperature.
The computed dust masses for a family of temperature distributions which
satisfy the previous constraints are comparable within the flux and dust
parameters uncertainties.

With $\alpha = 1$ and the fluxes given in Table~\ref{ngc128}, we
estimate the dust mass associated to the galaxies NCG~128 and NGC~127, to
be in the range $6\times10^6 M_{\sun} \div 1\times10^7 M_{\sun}$.  This value
is larger than those obtained by the single temperature models, because the
temperature distribution accounts for the contribution of the colder dust.

It has to be noticed that, since our constraint on the peak temperature is 
derived from the 60-100 \micron\ data, the computed mass may be biased by the
presence of the warm dust even if the temperature distribution estimates the
different contributions. In particular, this happens if a significant
fraction of cold dust is present in the source. 
In this case, we can consider our dust mass evaluation a lower limit for the
dust content of NGC~128 and NGC~127. On the other hand, the NIR dust mass
belongs entirely to NGC~128, and must be considered an upper limit, since it
was derived using the $R_V$ galactic value, i.e. the absorption of a spiral
galaxy for the early-type NGC~128.

Taking into account the high uncertainty of the NIR dust mass ($\sim 40\%$),
we thus suggest to adopt the dust mass of $6-10\times10^6$ M$_{\sun}$ for the
pair of galaxies NGC~128 and NGC~127, and a dust mass lower limit of
$\sim6\times10^6$ M$_{\sun}$ for NGC~128.

\section{Conclusions\label{Discuss}}
The large number of observational data presented in this work
can be summarized as follow:\\ 
1) the peanut shape of the galaxy is clearly visible at optical and NIR
wavelengths;\\
2) the color of the inner regions is rather uniform and similar to that of
the disk component. A small color gradient toward the center is observed;\\
3) the stellar disk is thick and distorted, in particular toward the NW side;\\
4) the galaxy host a counter-rotating gas component which
is tilted in the direction of the companion galaxy NGC~127. The mass of such
component is approximately $\sim 2.7\times10^4
M_{\sun}$;\\
5) the gas has not the same distribution of the dust component which is
largely confined in the region of interaction between NGC~128 and NGC~127.
The mass of the dust component is estimated of $\sim 6\times10^6 M_{\sun}$;\\
6) the gas does not seem  to fuel a nuclear activity or a starburst of star
formation;\\
7) the velocity field of the stellar component is approximately cylindrical.
The velocity dispersion along the minor axis is rather constant and high.
The RC of the gas component has the same slope of the stellar component.\\
8) the ``figure-of-eight'' of the RC, signature characteristic for the
presence of a bar, is not observed, either in the stellar and gaseous component.

The whole data-set presented in this paper does not clearly reveal or dismiss
the presence of a bar in NGC~128 and seem to support the theoretical mixed
scenario proposed by Mihos \etal\ (\cite{mihos:etal}). Interactions and 
soft merging events probably triggered a disk instability which originate a
strong buckling phenomenon. We can suspect that the interaction between
NGC~128 and NGC~125 or the accretion of a small gas-rich satellite was
responsible of such instability. The interaction with the spiral companion
NGC~127 seems very recent since dust and gas have not the same spatial
distribution in the galaxy.

\begin{acknowledgements}
We thank P. Rafanelli for the large field image of NGC~128 taken at Calar Alto,
R. Assendorp for the valuable help in the IRAS data reductions, and B. Jarvis
for the long slit RCs of NGC~128 perpendicular and parallel to the major axis.
We are also very grateful to the observers of the Spectial Astrophysical
Observarory, S.N. Dodonov, S.V. Drabek and V.V. Vlasiuk assisting us
at the 6m telescope, as well to J.L. Gach of Marseille Observatory for
his help during observations and data reduction.
\end{acknowledgements}

\appendix
\section{Long slit spectroscopy\label{SPM}}
NGC~128 and the template stars \object{HR~459} (K2III), \object{HR~6884}
(K0III), and \object{HR~8610} (K2II), have been observed with the 3.6m ESO
telescope equipped with EFOSC in three different nights of June 1990.

The detector was a RCA CCD $640\times1024$ pixels, used in binned mode. The
pixel size of 30 \micron\ maps $0\farcs68$ on the sky.  The grism gives a
dispersion of $\sim2$ Ang/pxl in the wavelength range 5000--7000.
The $3\farcm6\times 1\farcs5$ spectrograph slit was aligned with the
major and minor axes of the galaxy.

In the first night we got an exposures of 1 hour along the major axis
in the South direction, placing 
the center of the galaxy at one end of the slit in order to cover as
much as possible the whole galaxy. In the second night we put the slit in the
North direction, but unfortunately the weather was
cloudy and with bad seeing ($1\farcs7 - 2\farcs1$).
The third night we got a single centred spectrum of the minor axis.

After the standard procedures of bias and dark subtraction and
flat-fielding of the raw data, we corrected the distortion pattern by
deriving a line-by-line wavelength calibration from the comparison spectra. 

The science spectra have been conservatively adaptive-filtered following the
procedures described by Richter \etal\ (\cite{rich:etal}). This technique
allows to obtain reliable rotation curves (RCs) and velocity dispersion (VD)
profiles extended up to $\sim 2$ mag fainter in surface brightness.

In the spectra are clearly visible the absorption features of the MgI triplet
($\lambda\sim 5200$) and the Na D-band around $\lambda\sim 5900$. 
The analysis of the spectra is based on the Fourier Correlation Quotient
method (FCQ), as developed by Bender (\cite{bender}). We essentially used the
same procedures described with more details in D'Onofrio \etal\ (\cite{donof2}).

Unfortunately the low S/N ratio and the unfavourable spectral resolution
hampered the acquisition of an accurate broadening function and we were not
able to check the kinematical signatures of eventual subcomponents.
We could only verify that the change of the template stars introduces no
differences in the final RCs and VD profiles.

In our long slit spectra of NGC~128 we recognized five emission lines, 
[NII] ($\lambda$6548), \ha\ ($\lambda$6563), [NII] ($\lambda$6583), [SII] 
($\lambda$6717) and [SII] ($\lambda$6731). The most intense one is [NII]
($\lambda$6583), while \ha\ is attenuated by the neighbouring absorption lines.

\section{CIGALE\label{CIG}}
CIGALE data have been acquired at the 6m telescope of SAO (Special 
Astrophysical Observatory, Russia) on October 27, 1995. The scanning 
Perot-Fabry interferometer, installed inside the pupil plane of a  
focal reducer which was attached to the F/4 prime focus of the telescope, 
gives a spectral resolution of 9000.

The detector was an intensified photon counting system (IPCS) with a 
pixel scale of $0.97''$. The free spectral range was 606 \kms. The IPCS
enables a rapid scan of the interferometer (each of the 32
channels were scanned typically in 18 s). 
The blocking interference filter was of 12 Ang, centered on the
mean recession velocity of the galaxy.
The total exposure was 16000 s, during poor weather transparency conditions.

Reduction of observational data (correction for phase shifting, night sky
emission substraction, velocity and monochromatic maps) were done using
standard methods through the ADHOC software (see Plana \& Boulesteix
\cite{plana:boul} for details).
The final data were beamed $5\times5$ pixels because of the poor observational
conditions.

\section{CCD broad band photometry\label{Phot}}
CCD images of NGC~128 and of a set of standard stars were taken in two nights
in September 1994 through the $B$ and $R$ filters
at the 1.5m ESO Danish telescope. The frame dimensions are of
$1025\times1022$ pixels, with a scale length of $0\farcs38$ per 
pixel on the sky. The exposure times were respectively of 40 and 20 minutes.
The seeing was approximately $1''$ FWHM.

After the standard make-up procedures the images have been calibrated using
both a set of standard stars observed the same night and the photoelectric
aperture photometry taken from the Longo \& de Vaucouleurs (\cite{longo1}, 
\cite{longo2}) catalogs. The standard stars were chosen from the set of
Landolt (\cite{landolt}).

The calibration through the standard stars uses the following equations:
\begin{eqnarray}
B &\!\!=\!\!& -2.5 \log(I_B) + h_B (B-R) + k_B\\ 
R &\!\!=\!\!& -2.5 \log(I_R) + h_R (B-R) + k_R
\end{eqnarray}
where $I_B$ and $I_R$ are the fluxes in adimensional units of the
stars obtained after the normalization of the images to the standard
exposure of 1 sec and the correction for the extinction of the
atmosphere, $h$ and $k$ are respectively the color term and the zero
point of the CCD for each night in both filters.  They give the
transformation from our photometry to the $B$ Johnson and $R$
Kron-Cousin systems.

We also built the growth curves by integrating the flux through circular
apertures in order to compute the total luminosity of the galaxy.
By indicating with
\begin{equation}
C(\rho) = \int_{S(\rho)}^{} I(x,y) dxdy
\end{equation}
the growth curve of the galaxy, where $S(\rho)$ is the area of the circles of
radius $\rho$ centered on the galaxy nucleus, we converted in magnitudes the
growth curves through the set of equations:
\begin{eqnarray}
m_B(\rho) &\!\!=\!\!&  -2.5\,\log C_B(\rho) + h_B (m_B(\rho)-m_R(\rho)) + k_B\\
m_R(\rho) &\!\!=\!\!&  -2.5\,\log C_R(\rho) + h_R (m_B(\rho)-m_R(\rho)) + k_R 
\end{eqnarray}
adopting the zero point and color terms derived from the standard stars.

Following de Vaucouleurs \etal\ (\cite{rc3}) we considered the
galactic extinction toward NGC~128 very small and applied no correction. 
The $R$ magnitude was corrected to
the Johnson system using the transformation given by Longo \& de
Vaucouleurs (\cite{longo2}). 

The sky surface brightness was estimated using the standard stars and
the photoelectric aperture photometry of the galaxy for both nights with an
internal accuracy of 0.05 \mga. We got $\mu_B = 22.70$, $\mu_R = 21.20$
(first night) and $\mu_B = 22.66$, $\mu_R = 21.00$ (second night).
The sky counts were measured at the corners of the frames.

\section{CCD narrow band photometry\label{alphaphot}}
Images of NGC~128 through the interference filters \ha\ and the nearby
continuum have been obtained at the 1.5m Danish telescope during the same ESO
run of the optical observations. The adopted filter has a width of 50 Ang
and takes into account the redshift of the galaxy. The width is large enough
to include the contribution of \ha\ and [NII] lines. 

After the procedures of bias, dark, flat-field correction, cosmic
rays cleaning, sky subtraction, normalization at 1 sec exposure, reduction
to a standard airmass, and alignment of the images, we faced with the
probelm of isolating the \ha\ contribution with respect to the continuum.
The problem is to find the constant factor $\alpha$ which equates the
level of the continuum of the two images (\ha\ and nearby continuum):
\begin{equation}
I_l(x,y) = \alpha I_c(x,y)
\end{equation}
where $I_l$ and $I_c$ give respectively the fluxes measured through the
lines and the continuum.

In order to derive $\alpha$ it is convenient to establish the region of
the frame where one can exclude the contribution of the emission lines.
This is the region far from the center of the galaxy.
The $\alpha$ coefficient could be obtained from the growth curves of the
two images because the gradient of these curves scale with the same $\alpha$
factor:
\begin{equation}
\partial_\rho \mu_l(\rho) = \alpha \partial_\rho \mu_c(\rho).
\end{equation}
By increasing the radius of integration the contribution of the
emission is rapidly vanishing. It is therefore sufficient to calculate
$\alpha$ for large values of $\rho$ by excluding the central region.

In order to calibrate the \ha\ flux we used the standard star L870-2
of the Oke (\cite{oke}) catalog that was observed the same night.  The flux
$F_*$ of this star at $\lambda$6654.29, corresponding to our recession
velocity of 4180 \kms, is $1.933\times10^{-14}\pm0.024\times10^{-14}$
{\rm erg sec}$^{-1}$ {\rm cm}$^{-2}$ Ang$^{-1}$.  By adopting the previous
procedure of analysis to the standard star, one obtains the ratio
between the standard and measured flux $S = F_*/I_* =
1.14\times10^{-15}\pm0.05\times10^{-15}$ erg cm$^{-2}$ ADU$^{-1}$, which
express the sensitivity of our system.  Then, by building the growth
curve of the ``pure'' \ha\ image of the galaxy we get a total flux
$F_g = 3.08\times10^{-14}\pm0.61\times10^{-14}$ {\rm erg sec}$^{-1}$
cm$^{-2}$, from which the resulting \ha\ $+$ [NII] luminosity turns
out of $2.18\pm0.41\times10^{40}$ {\rm erg sec}$^{-1}$, a value comparable to
that of other S0 galaxies of the same luminosity 
(Buson \etal\ \cite{bus:etal})\footnote{The error on the total luminosity does
not take into account the error on the adopted distance of the galaxy.}.

\section{NIR photometry\label{Nirphot}}
Near infrared $JHK$ images of NGC~128 and of a set of standard stars were
secured at the Tirgo telescope with the Arnica NICMOS3 camera in October 1995.
The scale-length on the detector is $1\farcs0$ per pixel for
a total field of view of $4' \times 4'$. The seeing was about
$1.5''$ FWHM and the quality of the sky was photometric.

A sequence of on-off exposures have been taken for each filter
on the sources and the nearby sky with exposures ranging from 5 to 12 sec. 
The telescope was dithered each time in different position so that the stellar
images do not appear in the same place.
The standard stars have been taken from Hunt \etal\ (\cite{hunt2}).

The images have been reduced following the Arnica package provided by Hunt
\etal\ (\cite{hunt1}) for the IRAF system. Bad pixels have been corrected, after
flat-fielding, by linear interpolation using a reference mask. The cosmic rays
and the residual bad pixels have been eliminated using IRAF.

The Arnica zero point of the night for each filter was derived through the
aperture photometry of the standard stars observed before and after the
target galaxy. We used the average atmospheric extinction coefficients
given by Hunt \etal\ (\cite{hunt2}). 
The mean level of the sky background was $\mu_J = 14.95\pm 0.10$,
$\mu_H = 13.24\pm 0.10$, and $\mu_K = 12.23\pm 0.10$ \mga.
The sky counts level was measured at the corner of the frames.

\section{Dust mass from NIR data\label{ODM}}
In order to evaluate the dust content from the NIR data, we use the color 
excess $E(B-\lambda)$. Since the
color excess is related to absorption, we estimated the dust grain
column density and the optical depth using the model of Cardelli \etal\ 
(\cite{card:etal}, hereafter CCM) which uses the mean extinction curve 

\begin{equation}\label{eq10}
\langle A(\lambda)/A(V)\rangle = a(\lambda) + b(\lambda)/R_V
\end{equation}
where $R_V = A(V)/E(B-V)$ is the only free parameter.  The
coefficients $a(\lambda)$ and $b(\lambda)$ were derived as in CCM.
Taking into account the relation between the color excess and the
absorption we can write 

\begin{equation}
E(B-\lambda) = A(V)\lbrack R_V^{-1} + 1 - F_{CCM}(\lambda) \rbrack,
\end{equation}
being $F_{CCM}(\lambda)$ the right hand term of eq.\ref{eq10}. 
Assuming $R_V=3.1$ and the derived value of
$F_{CCM}(\lambda)$, we got $A(V)$. 
We estimated a visual absorption $A_V \simeq 0.3$ mag. 
Since $A_V$ is proportional to the optical depth, a dust grain
model has to be introduced to derive both the dust column density and
the dust mass:

\begin{equation}
M_d = {{4\over 3} {{a \rho_d}\over {Q_\lambda}}
{A_\lambda\over 1.086} A_{Gal}},
\end{equation}
a value which depends on the dust grains radius $a$, on the grain density
$\rho_d$, on the extinction efficiency $Q_\lambda$, on the absorption
$A_\lambda$ and on the area of the galactic region interested by the
absorption features $A_{Gal}$.  

Using the composite dust grain model of Mathis \& Whiffen (\cite{math:whi})
the dust mass turns out to be $6\times10^6$ M$_{\sun}$. It
is the dust mass excess associated to the North semiaxis.

\listofobjects
\end{document}